# Interleaving Lattice for the APS Linac

S. Shin,[1,2] Y. Sun,[1] J. Dooling,[1] M. Borland,[1] and A. Zholents[1]

[1] Argonne National Laboratory, Argonne, IL, 60439, USA
[2] Pohang Accelerator Laboratory, POSTECH, Pohang, 790-784, Kyungbuk, KOREA

E-mail: tlssh@postech.ac.kr

**Abstract**. To realize and test advanced accelerator concepts and hardware, a beamline is being reconfigured in the Linac Extension Area (LEA) of APS linac. A photo-cathode RF gun installed at the beginning of the APS linac will provide a low emittance electron beam into the LEA beamline. The thermionic RF gun beam for the APS storage ring, and the photo-cathode RF gun beam for LEA beamline will be accelerated through the linac in an interleaved fashion. In this paper, the design studies for interleaving lattice realization in APS linac is described with initial experiment result.

## 1. Introduction

New acceleration technologies for compact accelerators, such as laser-driven plasma wakefield accelerators, beam-driven plasma wakefield accelerators and structure-based wakefield accelerator (SBWAs), are rapidly developing many synergies. Development of laser plasma accelerators has made tremendous progress in generating soft x-ray spontaneous undulator radiation and in pursuing development of compact free electron lasers (FELs). SBWAs or plasma wakefield accelerators with SRF injectors are good candidates for multi-user compact FELs with a high repetition rate [1].

A critical requirement to realize SBWAs is to extract maximum power up to 80 % from drive bunches, and to obtain the highest energy for the witness bunch. A total of four times energy gain can typically be gained by witness electrons in SBWA, because the transformer ratio is five in the case of triangle distribution, assuming 80 % efficiency. Some activities of SBWAs for compact X-ray FELs are being conducted in ANL [2] and the development and test of a 0.5 m long acceleration unit are planned in the Linac Extension Area (LEA) tunnel downstream of the APS injector linac [3].

The APS linac [4, 5] is part of the injector complex of the APS storage ring. The thermionic RF electron gun (RG) provides the electron beam that is accelerated through the linac, injected into the Particle Accumulator Ring (PAR), cooled, and transferred to the Booster synchrotron. It is then accelerated, and injected into the APS storage ring. The linac is also equipped with a state-of-the-art S-band Photo-Cathode Gun (PCG). Three fast-switching dipole magnets at the end of the linac (so-called interleaving dipoles) direct the electron beam in and out of the PAR and into the Booster. Turning them off allows the beam to bypass the PAR and Booster and direct electrons into the LEA tunnel that follows the APS linac. The LEA beamline is being configured for the testing of small-aperture apparatus and other beam physics experiments that will take advantage of the high brightness beam generated by the PCG.

Typically, the beam generated from RG is used ~ 20 seconds every two minutes to support storage ring top-up operation. A relatively quick switching between RG and PCG (interleaving) will allow to operate LEA beamline during the rest of the two minutes. Because of the significantly different properties of beams produced by RG and PCG including beam energy, energy spread, bunch length, emittance and bunch charge, setting up one single lattice for the linac suitable for both beams is extremely challenging. In this paper we present a solution of this problem. Section 2 introduces optimum APS linac lattice for storage ring injection. Section 3 describes the optimum APS linac lattice for PCG and LEA experiments. A compromise interleaving lattice that conserves the high brightness of the PCG generated beam and maintains the high injection efficiency to PAR of the RG generated beam is described with experimental result in Section 4. Section 5 presents conclusions.

## 2. APS linac for beam injection

The APS injector linac (Fig. 1) includes twelve S-band accelerating structures and an asymmetrical magnetic chicane that contains four dipoles and two quadrupoles [6]. A thermionic RF gun and an alpha magnet are used to inject a short train up to 17 electron micro-bunches into the linac. (A hot spare set of the thermionic RF gun and alpha magnet is also installed). The PCG is installed at the front end of the linac and equipped with one additional accelerating structure upstream of the two thermionic RF guns. The chicane is not used for the compression of the RG generated beam as it is done by the alpha magnet, so the chicane is dedicated to compression of the PCG generated beam. Table 1 shows the main beam parameters.

The thermionic RF gun system includes the gun itself, the transport line from the gun to the entrance of the linac section, a fast kicker and alpha magnet. This system prepares an electron beam for injection into the APS storage ring. A tungsten dispenser cathode with a diameter of 6 mm produces an electron beam with up to 4.5 MeV beam energy and up to 1.3A peak macro-pulse currents. A fast kicker in RG gun system limits the total current injected into the linac. An alpha magnet injects the beam from the RF gun into a chain of S-band accelerating structures. The alpha magnet also serves as a bunch compressor equipped with a low-energy particle scraper. The transport line consists of four quadrupoles before the alpha magnet and three quadrupoles after it.

An asymmetric chicane with the time-of-flight parameter $R_{56}$=-65 mm was originally designed to increase the peak current delivered to LEUTL [5], and has been operated for the RG2 beam without bunch compression. To reduce emittance growth caused by CSR effect, an asymmetric configuration of the chicane and a small horizontal beta function at the exit of the last dipole magnet is used. A configuration with $R_{56}$=-65 mm is less sensitive to difficult-to-control timing and phase errors [6]. Then the linac accelerating sections L2, L4 and L5 are adjusted to minimize the beam energy spread and obtain the desired final energy for beam injection into the PAR. The beam is typically accelerated on crest of the RF phase in L2 because this beam is already compressed in the alpha magnet and satisfy the bunch length requirement for the beam injection into PAR. The dispersion function after the asymmetric chicane is closed by using two quadrupoles within the chicane system.

Since the requirement for injection into the PAR was satisfied, the APS linear accelerator has been operated with minimal beam loss through the linac after filtering low-energy electrons in the alpha magnet. To prevent beam loss, beta functions though the linac are kept $\leq$ 25 m. Two dispersion bumps (one in the chicane and one in the beam transport line from the linac to PAR) are present in the horizontal plane. Due to large dispersion function and beta function in the region before the PAR, the trajectory is carefully controlled in this region to prevent beam loss. Lattice functions at the end point are matched to those at the PAR injection point.

## 3. APS linac for advanced accelerator R&D

A LCLS type photo-cathode gun [7] has been installed at the beginning of APS linac to provide a low-emittance electron beam into the LEA beamline. The electrons are driven by a picosecond Nd:glass laser. The maximum field gradient in the gun cavity is 120 MV/m. High brightness electron beam with peak current up to 1 kA is prepared by first accelerating the electron beam to approximately 40-45 MeV in the first accelerating section (L1), then by creating the energy chirp in the second accelerating section (L2) and finally by compressing the bunch in an asymmetric chicane. Accelerator sections (L4 and L5) located after the chicane are used to minimize the electron beam energy spread and to obtain the desired final beam energy.

The goal of a new beamline in the LEA tunnel of the APS is to deliver a high-brightness electron beam from the APS injector linac and the photocathode RF electron gun into the experimental area in the LEA tunnel for beam physics experiments. Two conditions are important for advanced accelerator

R&D: (1) the beamline optics must be tunable and capable of producing round and flat beams at the center of the experimental area (IP) where the device under the test (DUT) will be installed; (2) drift space in the experimental area must be sufficient to accommodate DUTs with up to 1m length. The beta-functions in the entire beamline should be kept less than ~ 60 m and the total length of the beamline in the LEA tunnel must be less than 15 m.

Another set of requirements is driven by a need to measure the effect of the DUT on the electron beam. Most of DUTs are expected to produce an energy chirp along the electron bunch, and to kick the electrons along the bunch with variable strength in the horizontal plane. The energy chirp will be measured by a magnetic spectrometer installed at the end of the beamline. To measure the kick, we will map it to a coordinate at the location of the diagnostic YAG screen. Therefore, the part of the LEA lattice after the DUT is designed so that it converts the horizontal angle at the IP into the horizontal coordinate at the end of the beamline, and produces a relatively small vertical beam size at the same point. As examples, we consider two lattice solutions of the spectrometer and straight line for a round beam (Fig. 2). The betatron phase advance from the IP to the YAG screen at the end of the beamline is close to $\pi/2$ in the horizontal plane, and is close to $\pi$ in the vertical plane. To measure the energy chirp with high energy resolution ($2 \times 10^{-4}$), a 40-m vertical betatron function at the entrance of bending magnet and 0.2-m vertical betatron function with 1.2-m dispersion function at the YAG screen are realized.

## 4. Interleaving lattice

### 4.1 Lattice design

Because of the significantly different properties (e.g., energy, energy spread, bunch length, emittance, bunch charge) of beams produced by RG and PCG, setting up a single linac lattice that is suitable for both beams is an extremely challenging task. In this section, we will introduce interleaving lattice design scheme. An emittance-measurement system is installed after the chicane (Fig. 1); this system consists of three beam-size measurement screens; the phase advance for each screen is matched to 60°. We use the location of this measurement system as the matching point for the interleaving lattice design, because Twiss functions can be also experimentally determined by using this measurement system. Longitudinal matching involves adjusting the phase and voltage of L2 for PCG beam to obtain the desired peak current and energy after the chicane, and to match with RG energy at the matching point. Then L4 and L5 are adjusted to minimize the energy spread and obtain the desired final energy for both beams. Energy gains for each beam (from RG and PCG) along APS linac (Fig. 3) both have the same beam energy after accelerating section L2 and the same setting for L4 and L5 to gain the energy up to 375 MeV. Following longitudinal matching, transverse matching is done for both beams.

To see how Twiss function for a beam can be diverged at a condition that has been optimized for a different beam, we investigate the change of the Twiss function after simply assuming that the beamline has been optimized for a different beam. During transformation along a beam transport line, the orientation and shape of the phase ellipse changes continuously but its area remains constant. In matrix formulation, the Twiss functions are

$$\begin{pmatrix} \beta \\ \alpha \\ \gamma \end{pmatrix} = \begin{pmatrix} m_{11}^2 & -2m_{12}m_{11} & m_{12}^2 \\ -m_{11}m_{21} & m_{11}m_{22}+m_{12}m_{21} & -m_{12}m_{22} \\ m_{21}^2 & -2m_{22}m_{21} & m_{22}^2 \end{pmatrix} \begin{pmatrix} \beta_0 \\ \alpha_0 \\ \gamma_0 \end{pmatrix} \qquad (1)$$

where $m$ is transfer a matrix element for the particle tracking:

$$\begin{pmatrix} x \\ x' \end{pmatrix} = \begin{pmatrix} m_{11} & m_{12} \\ m_{21} & m_{22} \end{pmatrix} \begin{pmatrix} x_0 \\ x'_0 \end{pmatrix} \qquad (2)$$

If we consider a quadrupole doublet with focusing relation $f_1 = -f_2 = f$ and center-to-center distance $d$ between two quadrupoles, transfer matrix elements are given by

$$M = \begin{pmatrix} m_{11} & m_{12} \\ m_{21} & m_{22} \end{pmatrix} = \begin{pmatrix} 1-\dfrac{d}{f} & d \\ -\dfrac{d}{f^2} & 1+\dfrac{d}{f} \end{pmatrix}, \quad \dfrac{1}{f} = \sqrt{k}\sin\sqrt{k}L \quad (3)$$

where $L$ is quadrupole length, $k = \dfrac{g}{B\rho}$ when focusing is horizontal, and

$$g = \dfrac{dB_x}{dy} = \dfrac{dB_y}{dx} = \dfrac{B_0}{a} \quad (4)$$

is the field gradient where $B_0$ is magnetic field at the pole tip ($r = a$). The Twiss function of a low-energy beam degrades dramatically through a quadrupole doublet that is focused appropriately for a high-energy beam (Fig. 4). Therefore, in interleaving lattice design the quadrupole through the linac must be set specifically for a low-energy beam to keep a small Twiss function.

Small Twiss functions for PCG beam should be satisfied at the exit of photo injector system (Fig. 5, starting point) to keep small Twiss functions for a high-energy beam through the linac, because the high-energy beam passes through a low quadrupole focusing system like drift space (Fig. 4). Therefore, some parameters (e.g., laser spot size, solenoid field, gun RF phase, field gradient for the first RF accelerator) in the photo-injector system are fully explored to achieve small emittance and Twiss functions (at starting point in Fig. 5) by using genetic optimization with Astra [8] simulation. Results of a genetic optimization [9] are shown in Fig. 6. Here a trade-off between the achievable emittance and the maximum values of the beta-functions is shown. (Twiss functions and emittance in x and y planes that are almost the same, because all components are cylindrically symmetric in this simulation.) For example, accepting the normalized emittance of 1.2 μm will allow to reduce beta functions up to 5 m by reducing the laser spot size on the cathode. However, when the emitting spot size on the cathode becomes < 0.25 mm, the space charge effects become too strong preventing beam emission of the desired bunch charge. Longitudinal space charge effect increases as beam size decreases, and initial particles in the tail part cannot overcome repulsive space charge force despite the high RF field in the gun cavity.

Figure 5 shows the layout up to the matching point and energy ratio between PCG beam and RG2 beam. The following sequence of steps is executed to achieve transverse matching:

(1) optimizing the lattice for the beam generated by the RG. This step sets the quadrupole for the low-energy beam based on the result of Fig. 4;

(2) setting these quadrupole values for the PCG beam with energy scaling;

(3) optimizing the lattice for the beam generated by the PCG. This step uses two dedicated quadrupoles shown (Fig. 5, red bars) with the goal of keeping the beta functions in the Linac low;

(4) optimizing the PCG beam using Q11 to minimize horizontal beta function at the end of chicane to reduce the CSR effect [6],

(5) matching the diagnostics condition for emittance measurement using the four quadrupoles (Fig. 5, "4Qs"; and

(6) re-tuning the RG beam with four dedicated quadrupoles shown (Fig. 5, black bars) to match to the same lattice functions after the chicane as in the case of PCG lattice.

Finally, the lattice functions for both RG and PCG beams become the same after the matching point, because the beam energy and lattice functions at matching point after chicane are the same. The lattice functions change at each step during the process of interleaving lattice design (Fig. 7). Figure 7 (d) and 7 (f) show the final beta-functions in the interleaving lattice for PCG and RG beams up to matching point, respectively. All constraint conditions for the matching are fully satisfied and lattice functions are each close to their dedicated optimal values.

### 4.2 Preliminary experiment study

To demonstrate interleaving lattice design, the preliminary experiment study was performed with limited control knobs for Twiss function and emittance from PCG. Laser spot size for PCG and dedicated quadrupoles on the first accelerator section will be available later. The main purpose in the experiment study is to realize 100 % beam transmission for the PCG beam, and the same Twiss function with the RG2 beam at the matching point. Dedicated quadrupoles (Fig. 5, red bar) and corrector for the PCG beam are adjusted to achieve the main goal.

Following acceleration in the linac and chicane, a three-screen emittance measurement system is used to both the emittance and Twiss parameters (Table 2). Emittance is determined by an rms fit to the beam distribution measured at the three screens. Optical filter is used to prevent signal saturation on the screens. With only dedicated quadrupoles used for the PCG beam, Twiss functions of the beam are close to those of the RG2 beam in the matching section. The same operation conditions are used for both beams after the emittance measurement system because both beams have similar energy and Twiss functions at emittance that point. Matching quadrupoles can be used to optimize emittance and Twiss functions of the PCG beam (Fig. 7d). Measured emittance of the PCG beam was larger than design value (Table 2), so it will be reduced if a gun laser with small spot size is used.

We have considered lattice functions for two beams, but the trajectory of each PCG beam can be significantly distorted in the corrector set values for the RG beam, because the two beams have different energies. Fortunately, the trajectory of the PCG beam can be acceptable without iteratively correcting both RG2 and PCG beams; transmission of RG2 and PCG beams are up to 90 % and 100 %, respectively. Normal transmission of the RG2 beam during operation is up to 90 %.

## 5. Conclusion

We have described the lattice design and preliminary experiment results for APS injector linac in support of interleaving of the thermionic RF electron gun and photo-cathode RF gun. Despite significant differences of the original beams, an excellent compromise was found in simulation and realized in a preliminary experimental study. This compromise allows efficient injection of the high-charge RG2 beam into the Particle Accumulator Ring, and efficient transport of the high-brightness PCG beam to the experimental beamline in the linac extension area for testing of small aperture apparatus and other beam-physics experiments. As future work, we will continue the interleaving lattice optimization with additional quadrupoles installed for PCG beam front-end beta-function control. Advanced accelerator R&D experiment will be conducted in LEA with the linac operating in interleaving mode.

**Acknowledgments**
We wish to thank M. Borland, L. Emery, V. Sajaev, and N. Sereno for providing helpful information and the many useful discussions. This research was supported by the U.S. Department of Energy, Office of Science, under Contract No. DE-AC02-06CH11357. This research was also supported by the Basic Science Research Program through the National Research Foundation of Korea (NRF-2015R1D1A1A01060049).

.

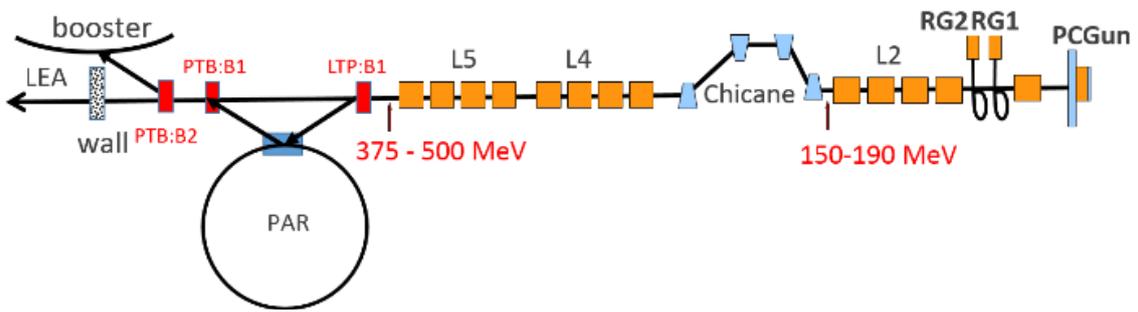

**Figure 1.** Layout of APS linear accelerator.

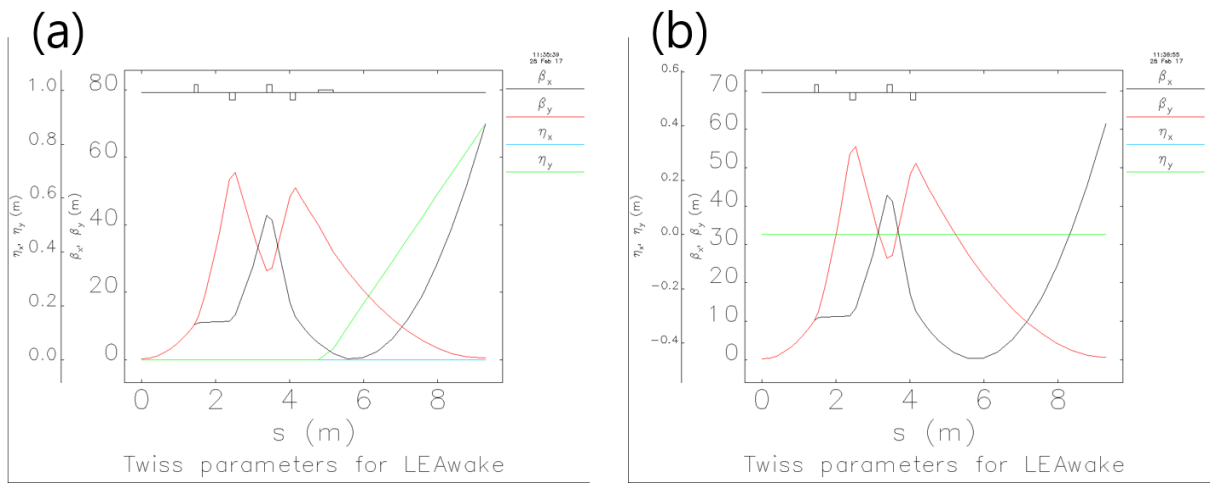

**Figure 2.** Lattice functions of spectrometer beamline and straight beamline in LEA for round beam.

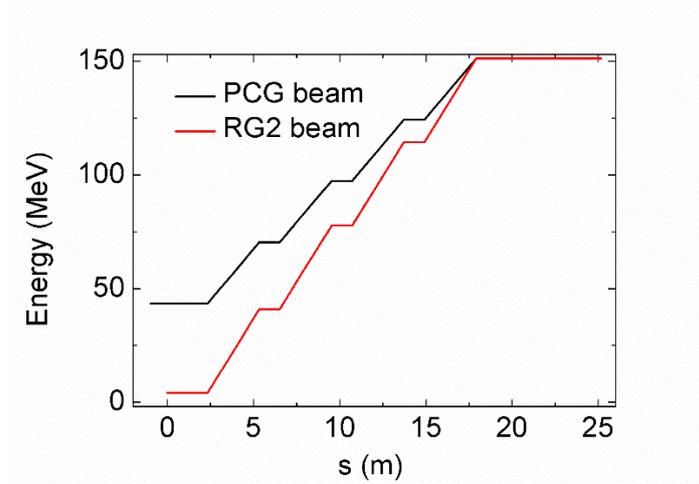

**Figure 3.** Energy gain along APS linac. Balck and red colors indicate RG and PCG beam, respectively.

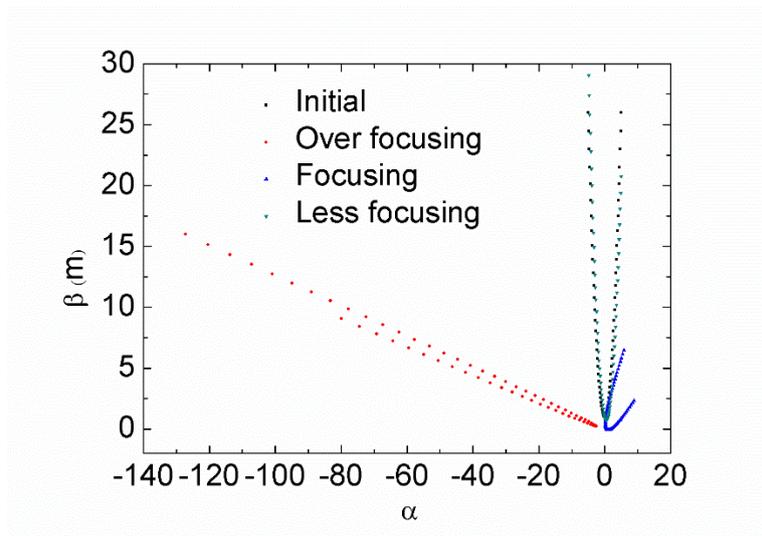

**Figure 4.** Twiss functions for initial, over focusing, focusing and less focusing cases. Here over focusing is corresponding to low energy beam focusing with quadrupole set for high energy beam.

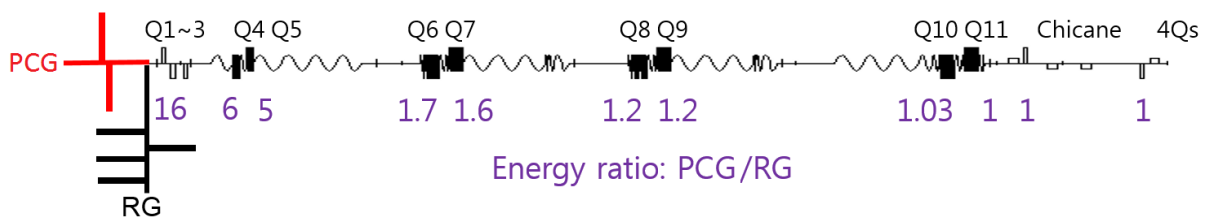

**Figure 5.** Beamline layout from each gun to the matching point downstream of the chicane. Energy ration between PCG and RG is also shown in the figure.

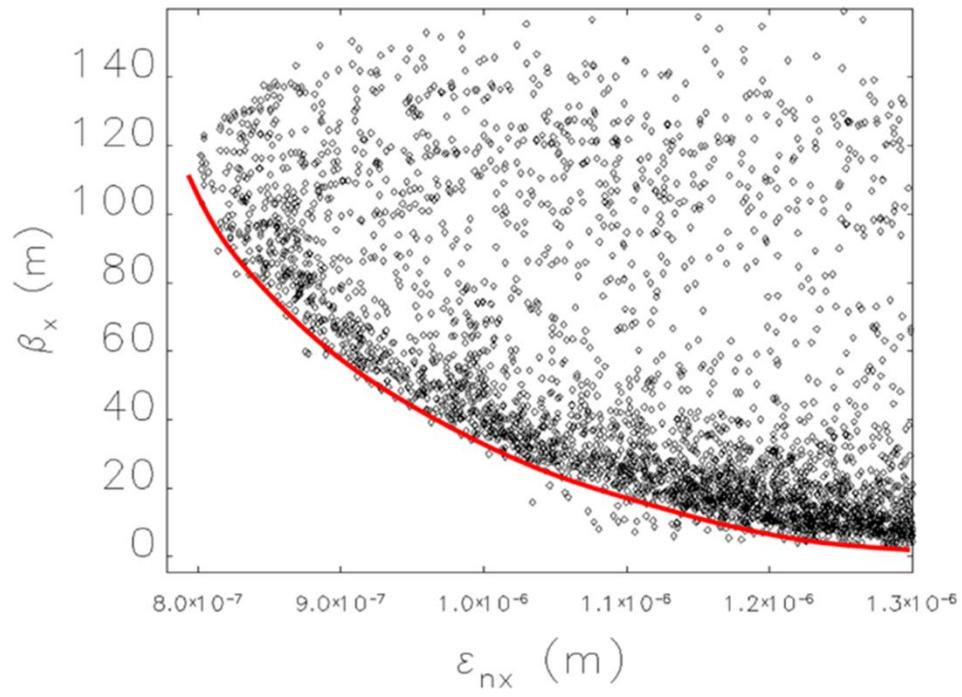

**Figure 6.** Results from MOGA: Beta functions vs normalized emittance.

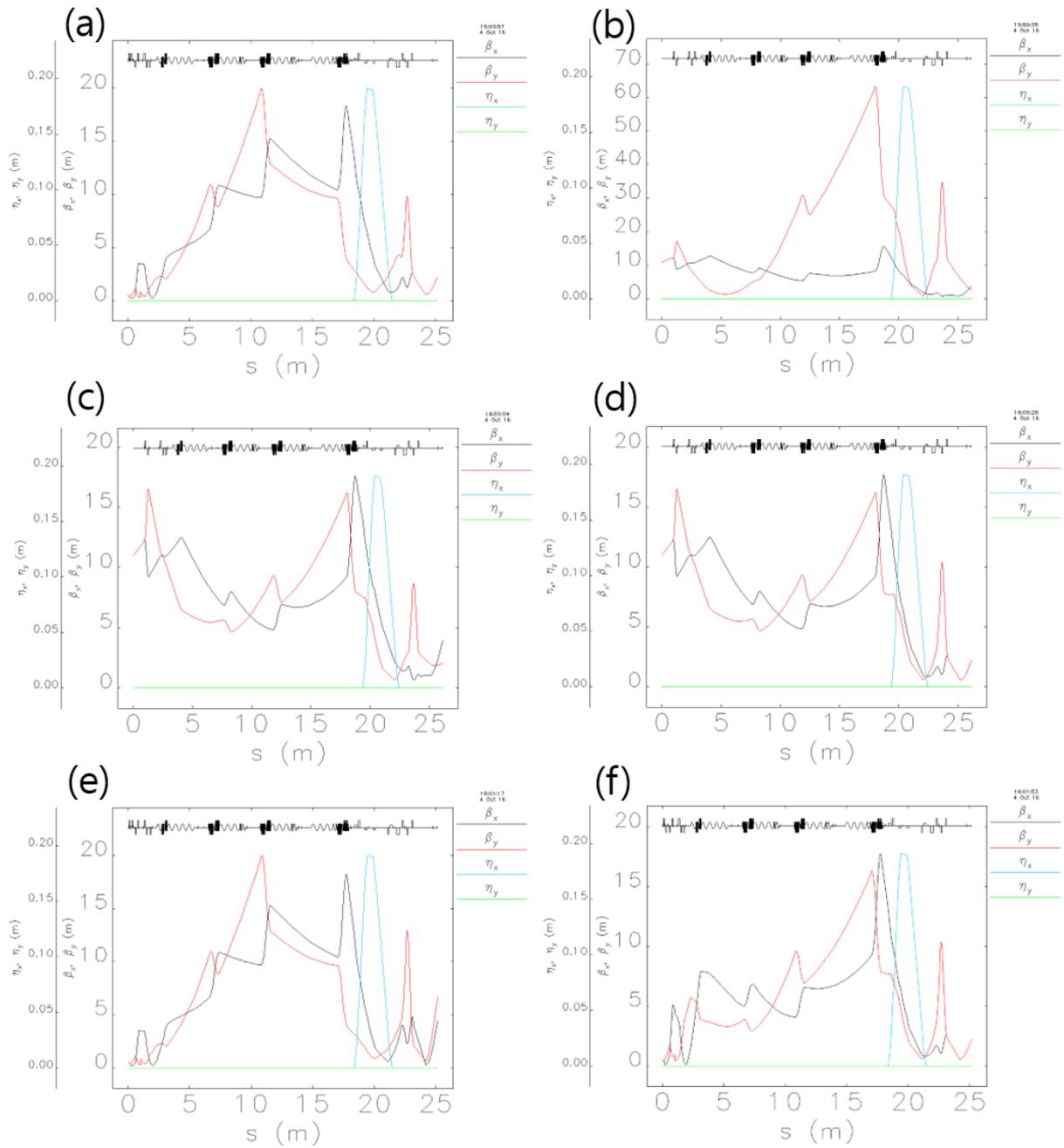

**Figure 7.** The results along the process of interleaving lattice design. (a) RG beam optimization. (b) PCG beam result by setting Quadrupoles obtained in (a) to PCG. (c) PCG optimization using two initial dedicated quadrupoles for PCG. (d) PCG optimization using Q11 and 4Qs for emittance measurement. (e) RG beam result by setting Q11 and 4Qs obtained in (d) to RG. (f) RG optimization using four initial dedicated quadrupoles for RG.

**Table 1.** Parameter for APS linear accelerator.

| Parameter | RG beam for Injection | PCG beam for R&D |
|---|---|---|
| Beam energy [MeV] | 375 | 375 |
| Beam emittance [um rad] | 15 | 1 |
| Charge [nC] | 1.6 | 0.3 |
| Peak current [kA] | 0.3 | > 1 |

**Table 2.** Result for emittance and Twiss function measurement.

| Parameter | $\beta_x / \beta_y$ | $\alpha_x / \alpha_y$ | $\varepsilon_x / \varepsilon_y$ |
|---|---|---|---|
| RG beam | 2.06 / 3.27 | 1.56 / 1.38 | 17.49 / 13.74 |
| PCG beam | 3.83 / 3.62 | 3.2 / 3.15 | 3.85 / 3.9 |
| PCG beam (matched) | 2.72 / 2.94 | 1.95 / 2.01 | 2.71 / 3.98 |